\documentclass[sigconf]{acmart}
\AtBeginDocument{%
  \providecommand\BibTeX{{%
    \normalfont B\kern-0.5em{\scshape i\kern-0.25em b}\kern-0.8em\TeX}}}

\usepackage[linesnumbered,ruled,vlined]{algorithm2e}
\usepackage{subcaption}
\usepackage{multirow}
\usepackage{graphicx}
\usepackage{booktabs}
\usepackage{xcolor}
\usepackage{enumitem}
\setlist{nolistsep}
\setcopyright{acmcopyright}
\copyrightyear{2023}
\acmYear{2023}
\acmConference[AIMLSystems 2023]{ The Third International Conference on Artificial Intelligence and Machine Learning Systems}{October 25--28, 2023}{Bangalore, India}

\begin{document}

\title[MobileASR]{MobileASR: A resource-aware on-device learning framework for user voice personalization applications on mobile phones}
\author{Zitha Sasindran, Harsha Yelchuri, T. V. Prabhakar, Pooja Rao}
\email{{zithas, harshay, tvprabs, poojarao}@iisc.ac.in}
\affiliation{%
  \institution{Indian Institute of Science}
  \city{Bangalore}
  \state{Karnataka}
  \country{India}
  \postcode{560012}
}


\renewcommand{\shortauthors}{Zitha Sasindran, Harsha Yelchuri, T. V. Prabhakar, Pooja Rao}

\begin{abstract}
We describe a comprehensive methodology for developing user-voice personalized automatic speech recognition (ASR) models by effectively training models on mobile phones, allowing user data and models to be stored and used locally. To achieve this, we propose a resource-aware sub-model-based training approach that considers the RAM, and battery capabilities of mobile phones. By considering the evaluation metric and resource constraints of the mobile phones, we are able to perform efficient training and halt the process accordingly. To simulate real users, we use speakers with various accents. The entire on-device training and evaluation framework was then tested on various mobile phones across brands. We show that fine-tuning the models and selecting the right hyperparameter values is a trade-off between the lowest achievable performance metric, on-device training time, and memory consumption. Overall, our methodology offers a comprehensive solution for developing personalized ASR models while leveraging the capabilities of mobile phones, and balancing the need for accuracy with resource constraints.
\end{abstract}






\maketitle
\section{Introduction and Motivation}
\label{sec:intro}

Over the past few years, we have witnessed a rapid improvement in  automatic speech recognition (ASR) tasks owing to the advancements in model architectures \cite{attention,rnnt_1,conformer,las,ds1}. Training these powerful models requires a significant amount of annotated and transcribed audio data. For example, Amazon Alexa\cite{alexa} trained their acoustic model with 1 Million hours of unlabelled speech and labeled speech of 7,000 hours. The tech giants, such as Microsoft, Google, Amazon, etc., use their efficient distributed training framework to train models in their cloud-based data centers. Data collected from various devices are offloaded to the cloud, trained using parallel machines in the data centers, and then the pre-trained models are downloaded on the devices for inference. There is a marked advantage of this paradigm as models become more general and robust.  However, the transfer of recorded speech to the cloud not only requires the device to be connected and requires a significant amount of internet bandwidth and, more importantly, results in privacy concerns. Furthermore, typically, the ASR models trained on generic datasets do not generalize well for users with different voice characteristics such as pitch, accent, and speaking rate. Hence, model adaptation\cite{speaker_adaptation1,speaker_adaptation2} or personalization \cite{disorderedspeech_personalization,investigation_personalization,ne_personalization} is crucial for better generalization of a user-specific ASR application. User voice personalization on such devices continues to be a challenge as recordings have to be trained on cloud-based models.

With the evolution of hardware and software technologies, the latest mobile phones are becoming increasingly powerful intelligent devices. This allows the researchers to bring machine intelligence from cloud-based data centers to mobile or edge devices. The idea is to preserve data privacy by keeping sensitive data on the users' device. Additionally, on-device personalization of models not only mitigates privacy risks but also enhances model performance by adapting to users' voices.  However, implementing such functionalities on mobile devices is limited by factors such as CPU speed, memory, and storage availability, as well as the quality of on-device training data due to the use of cheap sensor hardware. As a result, training end-to-end ASR models on lightweight systems is a challenge in itself.

A wide range of real-world problems can benefit from on-device training of ASR models on mobile phones. For instance, it can facilitate the adaptation of the user's voice for voice-controlled home automation or assistive technologies for individuals with speech impairments \cite{disorderedspeech_personalization}. 
The idea of resource-aware on-device training is critical in enabling the training of models on devices with limited computational resources. To address this issue, we propose a resource-aware sub-model-based approach for training ASR models on mobile phones, that takes into account the available resources on the devices. This paper provides a comprehensive overview of our resource-aware on-device training methodology for ASR models. We conduct experiments which consider device specifications such as RAM, CPU utilization, and training time across various brands of mobile phones. 
We present our findings from training the model with various accents on the device over a baseline model to mimic real-life scenarios.
Our findings also include the impact of tuning training parameters on the accuracy of the ASR model. Another important aspect of the training process is determining the stopping criteria as indiscriminate continuation can lead to overfitting and corruption of the model weights. We also incorporated the available battery percentage and the lower memory threshold in the mobile phones to make the decision. In summary, our approach offers a practical solution for effectively training ASR models on mobile devices.


We summarize our contributions through this work  as follows:
\begin{itemize}
    \item We propose a resource-aware sub-model-based training approach for ASR models that considers the storage, and battery capabilities of mobile phones. 
    \item We explore the correlation between available resources and training time, and demonstrate the efficacy of utilising sub-models for training in scenarios with limited resources.
    \item We conduct training by considering the evaluation metric, battery and memory constraints of the mobile phones and halt training accordingly.
    \item We demonstrate the working of our approach by deploying on multiple mobile phones and by using various accented speech to mimic real-life scenarios.
    \item We provide a complete overview of creating personalized models by considering multiple rounds of training.
\end{itemize}

This paper is organized as follows. In Section \ref{sec:related_works}, we first introduce the currently available techniques for on-device personalization. In Section \ref{sec:methodology}, we propose our resource-based adaptive on-device training procedure to train an ASR model on mobile phones. We then explain the implementation and experimental setup in Section \ref{sec:implementation}. In Section \ref{sec:experimental results}, we present the comprehensive results and discussions. We then present the summary and future scope in Section \ref{sec:summary}.

\section{Related works}
\label{sec:related_works}
Training an ASR model on resource-constrained personal devices such as mobile phones is a challenge, due to the limited memory, and compute capabilities of the devices. In a study by the authors of \cite{disorderedspeech_personalization,ne_personalization}, on-device personalization for shorter conversations. However, addressing longer conversations is challenging as the adaptation task will likely come with increased training times.
In  \cite{investigation_personalization}, the authors reduce the memory consumption during the on-device training by splitting the gradient computation into parts. The results demonstrated a  significant memory reduction at the expense of an increase in training time, which is not suitable for real-time deployments. In \cite{mdldroidlite,lowrank}, the authors introduced an on-device structure learning framework that enables resource-efficient deep neural networks on mobile devices. These works are limited to the simple tasks like speech command recognition, image classification, personal mobile sensing applications, and have not discussed about deploying a complex ASR model which uses longer utterances on the edge devices.  In \cite{rnnt_1,rnnt2} converts a complete pre-trained recurrent neural net model to an 8-bit integer quantized format to minimize memory consumption while training. It is important to note that most of these approaches primarily focus on simulation environments, and the complete deployment constraints of such methods on large-scale real-world mobile phones have not been extensively reported. Furthermore, none of these studies consider the resource capabilities of the devices during the model training process, which is crucial for real-time deployments.
 
In recent studies \cite{BN1}, a novel proposal suggests updating only the parameters of batch normalization (BN) layers. 
\cite{tinytl} reports the memory saving by reducing the activations by training only the biases in the model, freezing the weights in the model. The accuracy achieved using these methods remains similar to that obtained by fine-tuning the entire network.
Partial fine-tuning of specific layers is another option, yet the selection of the optimal number of layers for this approach remains arbitrary.
Several strategies such as sub-model-based training have been proposed to reduce the model size during the process of  client training. For instance, in \cite{fed_dropout}, the authors introduced a technique called federated dropout that randomly extracts sub-models to achieve the same goal. Additionally, \cite{heteroFL} and \cite{fjord} proposed a static model extraction approach in which a smaller sub-model is trained on the clients. Another work, \cite{fedrolex}, used a rolling window method to ensure that all parts of the model are trained at least once.
However, it is important to note that none of these studies have demonstrated real-time implementations for edge devices. Furthermore, the model selection approaches employed in these works do not consider the resources available on the devices. This aspect is crucial as the sub-model selection process should be informed by the capabilities and limitations of the edge devices.


\section{Resource-aware adaptive on-device personalization Methodology}
\label{sec:methodology}

While on-device training implementations in mobile devices for simple tasks such as image classification are common, no implementations exists that utilize long speech utterances and ASR models. Apart from the previously stated challenges, implementing on-device voice personalization for ASR tasks are amplified by larger model sizes and dynamic speech signal sizes. Our objective is to provide a comprehensive setup for on-device ASR personalization, with the ability to adaptively train complete models or sub-models based on the available resources on mobile devices.
We detail the methodology used to construct the on-device personalization task in this section. Our baseline acoustic model uses connectionist temporal classification (CTC)\cite{ctc} loss, but different model architectures can follow a similar approach.

\begin{figure*}[!ht]
	\centering
	\includegraphics[width =0.75\textwidth, height = 6.5cm]{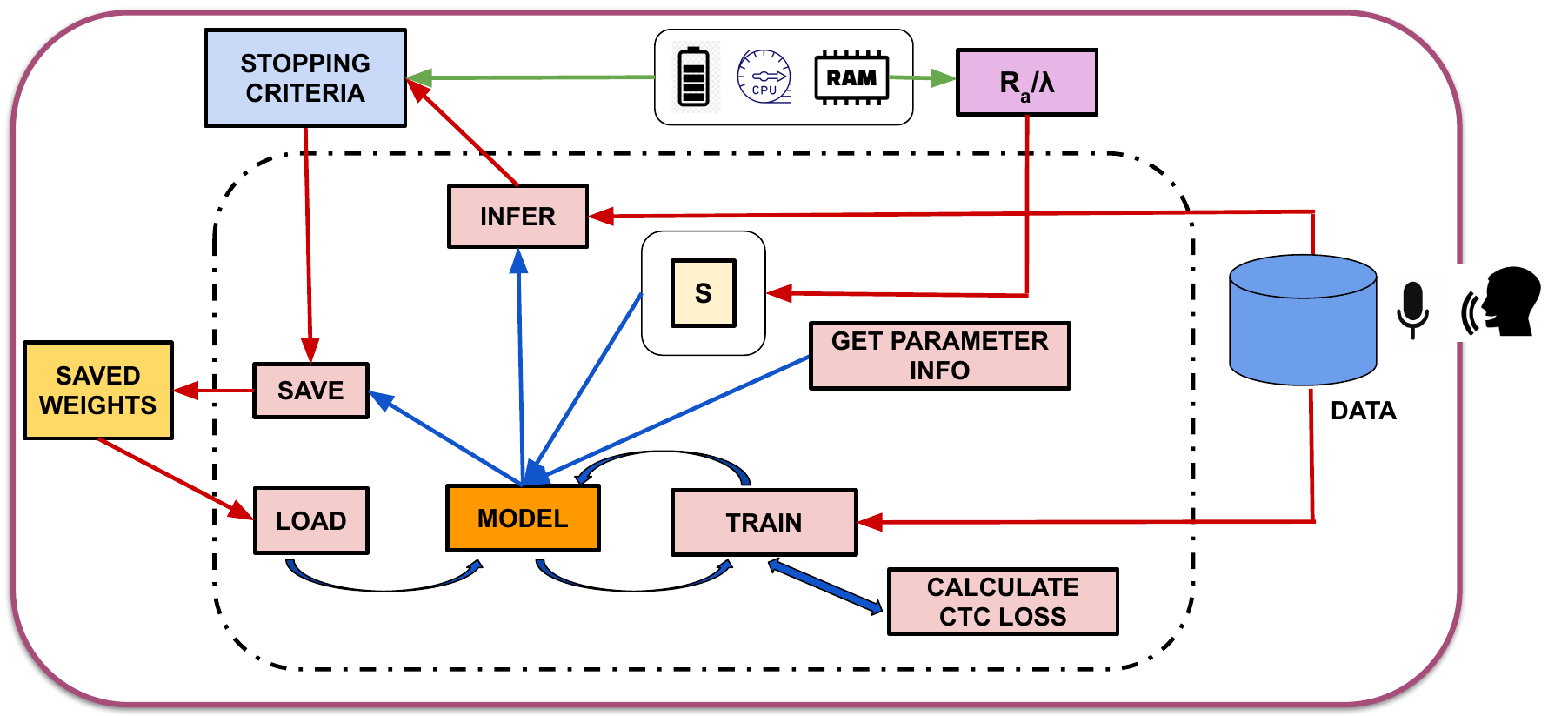}
	\caption{Resource-aware On-device training workflow: Optimized model with functions such as save, load, infer, train, calculate CTC loss and get parameter info, sub-model selection strategy using the ratio of available RAM ($R_a$) and the threshold ($\lambda$). $\mathbf{S}$ denotes the sub-model chosen for training. (Red and blue arrows denote interactions between the optimized model and external functions and internal functions, respectively. Green arrows denote interactions with system resources.) }
	\label{fig:workflow_training}
\end{figure*}

\subsection{Model for training} 
\label{sec:opt_model}
We need to convert the baseline ASR model to a memory-efficient and optimized format suitable for mobile platforms, with support for training the model. To achieve this, we implemented a conversion process using the Tensorflow \cite{tf} platform that utilizes a well-optimized Flatbuffer format for size reduction of larger trained models, with 32-bit floating point precision to represent the model parameters.
This optimized model facilitates faster inference, significant model size reduction and also enables training the model.
We use multiple tensorflow functions to interface with the model in flatbuffer format, and to specify the inputs (e.g., training data, inference data, path to the checkpoint file) and outputs (e.g., loss values, output probability matrix) for the optimized model as shown in Figure \ref{fig:workflow_training}. These functions can be customized based on our requirement. For instance, we built a function to extract the CTC loss value. Also, we use a function to get the parameter information from the model, later to use in our resource-efficient model selection approach. The train function takes a batch of mini-batch of inputs, and effectively train the model on device. 
In a conventional on-device training setup, four primary functions are employed to perform crucial tasks such as model training, output prediction, weight saving, and weight loading.







\begin{algorithm}[!t]
\SetKwInput{KwInput}{Input}
\DontPrintSemicolon

\SetKwFunction{FGetTrainingMode}{GetTrainingMode}

\SetKwProg{Fn}{Function}{:}{}
\Fn{\FGetTrainingMode{$R_a$, $\lambda$}}{
$\boldsymbol{\Theta}_S:\text{the trainable parameters obtained by selecting a }$\;
$\text{sub-model from the model } \mathbf{f}(\boldsymbol{\Theta}) $\;


\For{$k \gets 1$ to $L$}
{
$\boldsymbol{\Theta}_S \gets \left[\boldsymbol{\theta_{k}},\boldsymbol{\theta_{k+1}},\cdots \boldsymbol{\theta_{L}}\right]$\;
$\lambda \gets Estimate\_size(\mathbf{f}(\boldsymbol{\Theta_S}))$\;
\uIf{
$\frac{R_a}{\lambda}\geq 1$\;}
{

$model \gets \mathbf{f}(\boldsymbol{\Theta}_S) $\;
$break$\;}
}
}
\KwRet $\textrm{model}$\;
\caption{Resource-based training model selection}
\label{alg:model_selection}
\end{algorithm}





\subsection{Resource-aware model selection}
\label{sec:RA_model_selection}

\noindent \textbf{Motivation: }
Training a model on mobile phones with limited RAM can be challenging. Limited memory may lead to frequent garbage collection events which may pause the training process, as the OS needs to continuously reclaim memory to accommodate memory needed.
In \cite{edfed}, the authors show that there is an increase in training time when the available memory is low.
Also, Android OS incorporates a memory management feature that prioritises a few processes and therefore might terminate a few background processes, \cite{killingapps} when the system's available memory is less than a threshold \cite{ androidstudio}(can vary from one device to another). This action can potentially disrupt the smooth operation of the phone.


To alleviate these challenges associated with model training on limited memory, it is advisable to adopt memory management strategies like model pruning, or partial model-based training. Hence memory-aware approaches that adaptively adjust the model size based on currently available memory can help improve training stability.


\noindent \textbf{Our approach: }
In our resource-aware approach, we leverage custom training functions that can train specific parts of the model based on the currently available resources on the device. Let $\mathbf{f}$ denote the nonlinear function that is defined by our acoustic model and is characterized by the set of parameters denoted by $\boldsymbol{\Theta}$, with $L$ hidden layers.
Hence, the parameters of the model are given as  $\boldsymbol{\Theta} = \left[\boldsymbol{\theta_1},\boldsymbol{\theta_2},\cdots \boldsymbol{\theta_L}\right]$. The decision on which sub-part $\mathbf{S}$ to train is made based on resource information collected before the start of the training process and the parameter $\lambda$, which represents the amount of memory needed to run the training of the selected model on the device. So, we take a ratio of available RAM ($R_a$) and $\lambda$. This ratio indicates whether we can train the selected model with the available memory or otherwise. If the ratio is less than one, we select the next sub-model by freezing layers progressively. We continue freezing the layers until the ratio becomes greater than or equal to one. One key thing to note is we start freezing the layers from a top-down approach as many existing works showed that the bottom layers of the models are the ones that get personalized to the users.
Algorithm \ref{alg:model_selection} contains the detailed steps involved in the model selection. 


\noindent \textbf{Estimating $\lambda$ parameter}
Setting a threshold with respect to available RAM is a crucial aspect of running a model efficiently and preventing memory-related issues. RAM is a finite resource, and deep learning models can be memory-intensive. When running a model, it's essential to ensure that it fits within the available RAM to avoid crashes, slowdowns, or out-of-memory errors.
Hence we set the threshold $\lambda$ by estimating the memory usage considering the different stages, such as loading weights ($R_{L}$), and training with back-propagation ($R_{T}$) of the sub-model selected.
\begin{align*}
    \lambda = R_{T}+ R_{L}
\end{align*}
$R_L$ is the estimated RAM usage while loading the weight tensors and related information from a checkpoint to the model before training commences.
$R_T$ includes both the memory required to store the model parameters (weights and biases) and the memory required for intermediate computations during forward and backward passes (activations, gradients, optimizer specific parameters etc). We use the ``get parameter info" function in the Figure \ref{fig:workflow_training}, to extract the number of parameters in each layer of the model from the flatbuffer format.

Overall, at each round, the choice of sub-model to be trained should be based on practical considerations such as computational resources and time constraints. 
We use this approach to deploy our model on mobile devices to efficiently train the sub-models based on the resource availability.

\begin{algorithm}[!t]
\SetKwInput{KwInput}{Input}
\DontPrintSemicolon

\KwInput{No. of epochs $T$, battery threshold $b$, memory threshold $m$,  patience parameter $p$}
 $R_t\gets \textrm{Get RAM information from the mobile phone}$\;

 $\textrm{model} \gets \textrm{GetTrainingModel}(R_t, \lambda)$\;

 $i \gets 0$, $\textrm{past}\_\textrm{WER} \gets 0$\;
 $\textrm{current}\_\textrm{patience} \gets 0$\, $\textrm{stopping}\_\textrm{criteria} \gets True$\;

\While{$(i<T)~\textrm{and}~(\textrm{stopping}\_\textrm{criteria})$}{ 
    $B_a, R_a \gets \textrm{Get battery, and available memory information}$\;
    
    \uIf{$(B_a \leq b)  ~or~ R_a<=m$}{
         $\textrm{stopping}\_\textrm{criteria} \gets \textrm{False}$
    }\Else{
         $\textrm{current}\_\textrm{WER} \gets \textrm{Train}(\textrm{model})$\;
        
        \uIf{$\textrm{current}\_\textrm{WER} \leq \textrm{past}\_\textrm{WER}$}{
             $\textrm{current}\_\textrm{patience} \gets \textrm{current}\_\textrm{patience} + 1$\;
            \uIf{$\textrm{current}\_\textrm{patience}\geq p$}{
                 $\textrm{stopping}\_\textrm{criteria}\gets \textrm{False}$\;
            }
        }\Else{
             $\textrm{current}\_\textrm{patience} \gets 0$\;
        }
    }
    
     $\textrm{past}\_\textrm{WER} \gets \textrm{current}\_\textrm{WER}$\;
     $i \gets i + 1$\;
}

\caption{Resource-aware on-device training process}
\label{alg:adaptive-odt}
\end{algorithm}



\subsection{Training procedure} 
\label{sec:training}

Due to limited resources available on mobile phones, on-device training has to make sure that sufficient resources are made available to train the model long enough to learn the mapping, but careful enough not to overfit the training data.
We propose Algorithm \ref{alg:adaptive-odt}, Resource-aware on-device training process, keeping these two constraints in check. The algorithm is as follows:
\begin{itemize}
  \item The training procedure begins after the sub-model has been determined and continues as long it satisfies the stopping criteria threshold. The stopping criteria is a combination of two thresholds:
  the first threshold is battery, and memory-based, while the second one is accuracy-based.
  \item The training procedure is halted either when mobile phones' current battery ($B_a$) degrades below a certain percentage (b \%), or if it reaches the lower memory threshold (m). 
  \item The accuracy threshold stops the training procedure if the model does not improve over the ``p" (patience parameter) number of epochs. 
\end{itemize}


Given the limited storage available on mobile phones, we propose an approach to handle the training data efficiently. We train the selected sub-model with N utterances and then remove the stored data upon completing one training session.  The next session is resumed after N new utterances are acquired.
In each session, we train the sub-model using the train signature for T epochs with a batch size of B. The checkpoints at the end of each epoch are saved and restored using save and load signature functions. The decision to save the on-device trained model or resume training is outlined by the stopping criteria as described previously. 
The personalized model saved at the end of training is then used to translate the newly saved recordings using the predict signature. 
Figure \ref{fig:workflow_training} shows the optimized model with the workflow for training. 

\begin{table}[!t]
	\centering
	\caption{Total number of trainable parameters, and WER for sub-models. The initial WER before training is 37.46\%.}
	\label{tab:sub_models}
	\begin{tabular}[t]{clcc}
		\hline
		\textbf{Name}&\textbf{Sub-models}&\textbf{No. of trainable}&\textbf{WER}\\
		&&\textbf{parameters} &\\
		\hline		
        S1 &CONV 1-FC 2&30.24M&27.37\\
        S2 &CONV 2-FC 2&30.23M & 28.08 \\
        S3 &CONV 3-FC 2& 30.21M &28.13  \\
        S4 &BLSTM 1-FC 2& 30.19M & 27.7 \\
        S5 &BLSTM 2-FC 2&  19.97M &28.27 \\
        S6 &BLSTM 3-FC 2&  13.67M & 29.72\\
        S7 &BLSTM 4-FC 2& 7.38M & 31.42 \\
        S8 &FC 1-FC 2&   1.08M &35.19\\

        \hline
  \end{tabular}
\end{table}%

\section{Experimental setting}
\label{sec:implementation}
This section details the experimental setup used for resource based on-device training approach deployed on mobile phones.

\subsection{Resource-aware on-device training}
\label{sec:RA_ODT_setup}

In this section, we will discuss both the baseline ASR model used and the model size compression techniques used to deploy on mobile devices.
For our experiments, we utilized a pre-trained end-to-end acoustic model with a DeepSpeech2 \cite{ds2} architecture as the acoustic model. Appendix \ref{app:model} and \ref{app:datasets} provides specifications regarding the model and datasets used.  During each training round, the sub-model extraction approach involves the selection of a specific sub-model from the available categories, as described in Section \ref{sec:RA_model_selection}. Table \ref{tab:sub_models} shows the various sub-models , along with the number of trainable parameters for each sub-model and WER when deployed on a Oneplus 7T phone with available RAM $R_a$=4.7GB, and trained for 4 epochs. We can infer that the final WER achieved after 4 epochs is declining with freezing of layers of the models.
Given the multitude of possible sub-models, we have chosen three representative models, namely $S_1$, $S_5$, and $S_8$, which cover distinct parameter ranges.

We determined the value of $\lambda$ for the sub-models by incorporating the parameters involved in the training process. As outlined in Section \ref{sec:RA_model_selection}, we initially estimated $R_T$ and $R_L$ for the model. To compute the memory footprint required for training, we considered the various stages of training while using the function ``train" in the Figure \ref{fig:workflow_training}, where parameters need to be stored in memory. These stages include loading the input and output ($n_i$), size of the model ($n_m$), loss calculation ($n_l$), storing gradients ($n_g$), activations ($n_a$), errors ($n_e$) and optimizer parameters ($n_o$). To find the total number of parameters to be stored, we summed up all these components appropriately, giving us the total size $N_T$ to be stored. Since all the parameters are in the 32-bit floating-point data type, we calculated the estimated memory required for training in giga bytes (GB) as $R_T = \frac{N_T * 4} {1024^3}$.
The memory requirement $R_L$ is for executing the restore function in Figure \ref{fig:workflow_training}. It is determined based on the model's parameter size ($n_m$) with additional overhead needed to store node names and data types of the weight tensors. Moreover, both the checkpoint and the model in flatbuffer format need to be loaded into RAM during this process.

\subsection{Dataset} 
\label{sec:data acquisition}
A critical step in an on-device personalization framework is to adapt the model to user-specific data \cite{domain_adaptation1, domain_adaptation2}.
The limited computational power, memory, and storage capacity of the mobile phones, can hinder the collection and processing of large amounts of audio data required for effective on-device personalization. Annotating and labeling the recorded speech data is a labor-intensive task
\cite{investigation_personalization,ne_personalization}. In our setup, we load pre-filled speech transcriptions for the user to create training dataset by recording N utterances with transcripts displayed on the mobile screen, chosen at random from the transcriptions. The utterances are saved in the storage of the mobile phone. The application supports the user to either choose to save the utterance into the storage or re-record based on the clarity of the recordings.  In our future work, we aim to address this challenge by incorporating pseudo labelling \cite{nst} or semi-supervised learning \cite{semisupervised} techniques to handle scenarios where clean labels are not readily available.

To simulate unique speakers with various accents, we use the audio corpus detailed in Appendix \ref{app:datasets}.
The validation set is created in such a way that a few of the words are derived from the root words presented in the training set. This explains whether the model can predict the similar words in the test set effectively, without overfitting to the data.
Furthermore, to evaluate our approach, we collected real-time recorded data from two  subjects (1 each from male and female voice) using the application.

\subsection{Mobile phone based deployment}

To demonstrate the on-device training for ASR models, we develop an Android application for mobile phones.  To extract the necessary resource information, we utilize the Android developers' tools for memory management and the Android battery manager. With the memory management tool, we can closely monitor memory usage of the ASR model throughout the training process. Meanwhile, the Android battery manager enables us to track the battery consumption of the mobile phone during training. The detailed description about the hardware specifications of the mobile phones used and Android application details are given in Appendix \ref{app:andoid_app}.
In our work, we deploy our resource-aware on-device training protocol aimed at improving speech recognition across multiple mobile devices. This provides an insight on deployment constraints such as training time, accuracy, and memory for different hardware specifications.


\subsection{Experiments}
We conducted exhaustive experiments to study the efficacy of personalization.  These experiments include training for multiple accents over multiple rounds. Also, we tested the approach with real-time recordings from a limited set of users.

\subsubsection{Hyperparameter tuning}
Our intent is to measure the effect of batch size on the model size,  RAM usage, and CPU utilization for on-device training. Given a model, we find the best possible configuration for training in a mobile environment. For this experiment, we randomly select one speaker and train the model using Adam \cite{adam} optimizer with learning rate of the model to 10e-5. 

Along with batch size, another critical tuning parameter for training a model on-device is learning rate, as it determines the convergence time and accuracy of results. We compare two learning rates, 10e-5 and 10e-6 for different sub-models. We chose a learning rate of 10e-5, as the pre-trained baseline model was trained with same. We compare it with a lower learning rate of 10e-6, as we are fine-tuning the model for a particular user. Hence, a lower learning rate might move towards the minimum and prevent overshooting.

\subsubsection{CPU and RAM utilization}
\label{sec:cpuandmemoryutilisationondifferentphones}

In this experiment, our intention is to study the CPU utilization and RAM usage during the training process on multiple phones.
We use the Android memory management and battery manager toolkit to measure the real-time data usage of our application after every second.



\subsubsection{Training multiple accents for multiple rounds}
We study the performance of model saved during multiple rounds of training, using WER based stopping criteria. We use the multiple accented-dataset detailed in Appendix \ref{app:datasets} for this experiment. We set $N=80$ for our experiments, with 60 samples for training and 20 samples for validation.

We implement a simple training pipeline to personalize the baseline model with user-specific data in two rounds. In round one, after the data acquisition, the model weights are updated till $T^{th}$ epoch. The best weights are saved at epoch $E$ based on their respective stopping criteria. This signals the end of one round. The stored data is deleted, and second round commences when the user inputs new $N$ utterances . For round two, first, we load the weights saved after round one to the model, then we extract the sub-model-based on the resource information, and resume training with the newly added dataset. Our intent is to study the generalization of the model trained with new data instances for each round. 

\subsubsection{Real-time recordings}

To show the efficacy of the on-device training approach, we use TTS generated samples which do not have external noise or other artifacts. However, it is actually important to validate the framework  when directly used by the end users. To this end, in our experiment, we took one mobile phone and requested the speakers to read out the transcripts displayed on the application's GUI screen as shown in Figure \ref{fig:recording}. We recorded the samples on the phones, in the same location, to ensure that the mobile phone capture data from similar recording environment.  Thus, we validate that the on-device training with real-time recordings are behaving as expected.

\section{Experimental Results}
\label{sec:experimental results}
In this section, we present the experimental results from our on-device training settings explained in Section \ref{sec:implementation}. By using the setup explained in Section \ref{sec:RA_ODT_setup}, we estimated $\lambda$ for the complete model as 2GB with $R_T$, and $R_L$ as 1.65GB and 325MB respectively. The detailed overview about the estimation of $\lambda$ for the selected model is given in Appendix \ref{app:memory_estimation}. We set the battery threshold $b$ as 25\%.

\begin{figure*}[!t]
    \centering
	\includegraphics[width = 14cm, height = 8cm]{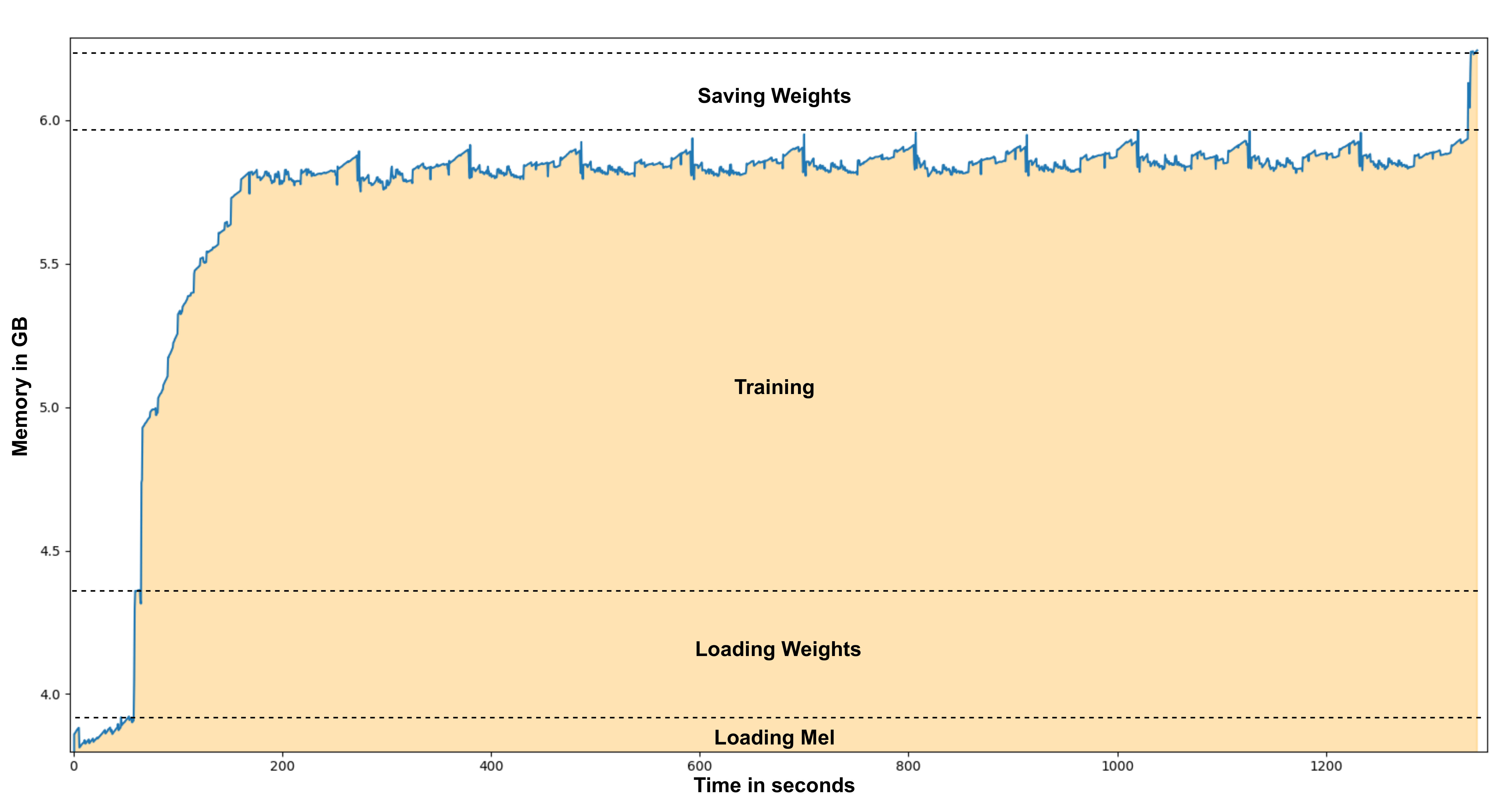}

	\caption{RAM utilization for OnePlus7T with batch size 5 with an initial available memory of 3.8 GB }
	\label{fig:cpuandramgraph}
\end{figure*}

\begin{table}[!ht]
	\caption{Time taken per epoch for training $S_1$ model in One Plus 7T for different batch sizes.}
	\begin{center}
		\begin{tabular}{c c c c c c}
			\hline
			\textbf{Batch}&\textbf{Opt. model}&\textbf{CPU}&\textbf{RAM}&\textbf{Time per epoch}\\
			\textbf{size} &\textbf{size in} & \textbf{(max)}&\textbf{(max)} &\textbf{Oneplus-7T}\\
			\textbf{(B)}&\textbf{(MB)} &\textbf{(in $\%$)} &\textbf{(in GB)} &\textbf{(mins)} \\
		\hline
			1 & 152.3 & 25 & 2.1& 80 \\
			2 & 153.3 & 25 & 2.3 & 51 \\           
			5 & 156.2 & 25 & 3 & 22 \\
			10 & 161 & 25 & 4.3 & 13 \\
			
		\hline
		\end{tabular}
	\end{center}
	\label{batch_size_vs_epoch_time}
\end{table}

\subsection{Optimal hyperparameters selection}
\subsubsection{Batch size}
The training was carried out on a OnePlus 7T phone with available RAM of 4.5GB, while the battery was being charged. Due to the ample resources, our resource-aware on-device training procedure, as described in Section \ref{sec:training}, selected the sub-model $S_1$ for this particular experiment. 
Table \ref{batch_size_vs_epoch_time} shows the time taken per epoch during training, the size of the model created for different batch sizes, the maximum CPU utilized, and RAM usage during training. We train our model with learning rate of 10e-5 and varying batch size.  As the batch size increases from 1 to 10, the size of the model increases from 152.3 to 161 MB. For the batch size of 1, the time taken to complete an epoch while training is nearly 80 minutes, and the memory used is 2.1 GB. As we increase the batch size to 10, the time per epoch decreases to 13 minutes, but the memory consumption increases to 4.3 GB. Therefore, as the batch size increases, the RAM consumption also increases to about 96\%. We selected an optimal batch size of 5 for our subsequent experiments. We conclude that selecting the right batch size is crucial to RAM utilization, and training time.

\begin{table}[!ht]
	\centering
	\caption{Comparison of different learning rate for sub-models on OnePlus 7T with different RAM conditions.}
	
	\label{learning_rate_comparison}
	\begin{tabular}[ht]{l c c c c}
		\hline
		\multicolumn{1}{c}{} & \multicolumn{2}{c}{\textbf{10e-6}} & \multicolumn{2}{c}{\textbf{10e-5}}\\
		\hline
		\textbf{Models}&\textbf{WER}&\textbf{Epochs}&\textbf{WER}&\textbf{Epochs}\\
		&\textbf{(\%)}&\textbf{(E)}&\textbf{(\%)}&\textbf{(E)}\\ 
		\hline
		Baseline &19.49&-&19.49&-\\
		\hline
		$S_1$ &13.4&12&14.2&1\\
		$S_5$ &13.4&13&13.4&2\\
		$S_8$&19.02&4&17.04&3\\
		\hline
	\end{tabular}
\end{table}%

\subsubsection{Learning rate}
In this experiment, we evaluated multiple sub-models on a One Plus 7T phone. We employed our training procedure and selected each sub-model by manipulating the available RAM on the phone through the addition of background apps.

Table \ref{learning_rate_comparison} compares the two learning rates of the different sub-models using the minimum WER and the number of training epochs required. The baseline model achieves a WER of 19.49\%. As expected, the minimum WER achieved by all models for each learning rate is lower than that of the baseline model. Therefore, we can infer that the model has learned from user-specific data for both learning rates. The learning rate of 10e-5 converges to least WER faster than 10e-6, across all models. The $S_1$ model converges to a WER of 13.4\% at 12th epoch with 10e-6 learning rate, whereas it converges to 14.2\% in one epoch with 10e-5 learning rate. $S_5$ model achieves the same least WER of 13.4\% for both learning rate. The downside is that 10e-6 learning rate takes nearly 10 more epochs compared to learning rate of 10e-5. The training time for $S_8$ model is shorter but the least WER achieved is nearly 5\% higher compared to other models.

\begin{figure*}[!t]
	\centering
	\begin{subfigure}[b]{0.45\textwidth}
		\centering
		\includegraphics[width = 1.0\textwidth,height = 4.5cm]{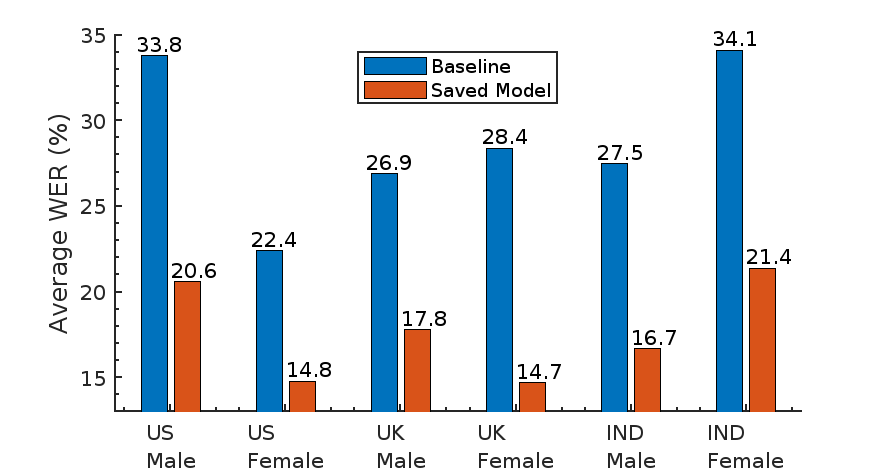}
		\caption{WER for multiple accents in round 1}
		\label{fig:wer_ne}
	\end{subfigure}	
	\begin{subfigure}[b]{0.45\textwidth}
		\centering
		\includegraphics[width = 1.0\textwidth]{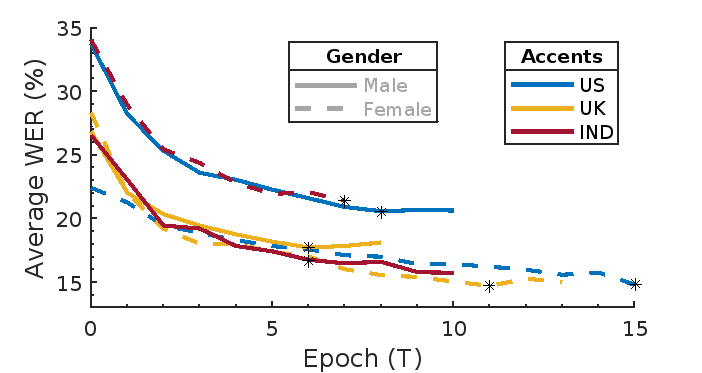}
		\caption{The WER trend across epochs for round 1}
		\label{fig:ne_wertrend}
	\end{subfigure}
	\caption{Round 1 results for multiple accents.}
  	\label{fig:NE_dataset results}
\end{figure*}

\begin{figure*}[!t]
	\centering
	\begin{subfigure}[b]{0.45\textwidth}
		\centering
		\includegraphics[width =1.0\textwidth,height = 4.5cm]{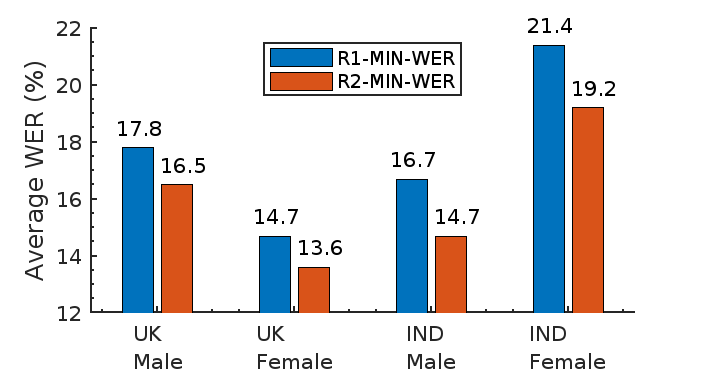}
		\caption{WER for multiple accents in round 2}
		\label{fig:r2werne}
	\end{subfigure}
	\begin{subfigure}[b]{0.45\textwidth}
		\centering
		\includegraphics[width =1.0\textwidth]{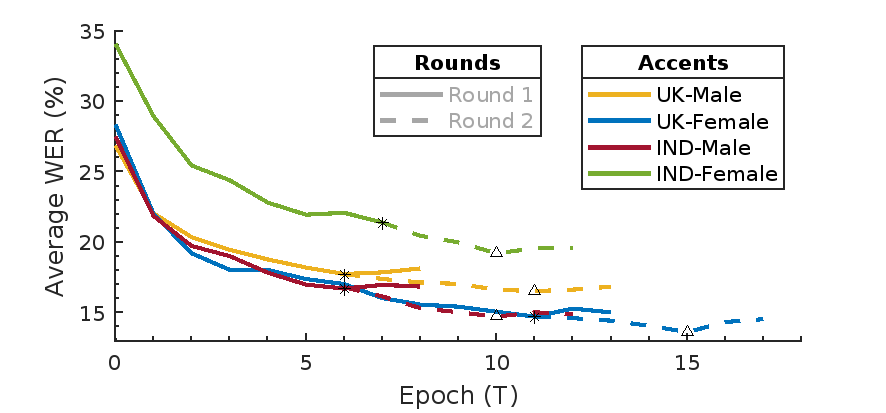}
		\caption{The WER trend across epochs for two training rounds}
		\label{fig:r2wertrend}
	\end{subfigure}
	\caption{Round 2 results for multiple accents}
	\label{fig:r2ne_bargraph}
\end{figure*}

\subsection{CPU and RAM utilization}
\label{cpuandmemory}

Our experiment focused on examining the impact of training on the CPU performance and memory of an android phone. Specifically, we utilized a OnePlus 7T phone that ran on Android 10 and had 4.5GB of available RAM out of 8GB. Our resource-aware model selection algorithm determined that the $S_1$ model was the best choice for our training purposes.

We use a learning rate of 10e-5 and batch size of 5. We train the model for one epoch on US Male voice, and measure the resources by sampling the data every 1 second during the various processes of on-device training and presented it in Figure \ref{fig:cpuandramgraph}.
From Figure \ref{fig:cpuandramgraph}, we observe several trends during the different stages of training process. 

In the pre-processing stage, the input mel-spectrogram features are extracted from an audio file. During this stage, we observe sharp spikes in CPU utilization from an average of 12\% to 25\%. The majority of the CPU is used by transformation functions in the feature extraction algorithm.
However, the memory consumption during pre-processing is very small (around 1MB for sample).
During the training process, the CPU utilisation is almost constant with an average of 12\% but with occasional spikes leading to a 15\% CPU usage.
However, memory consumption is high during the training process. It can be divided into 3 different sub-phases. Before the start of training, the optimized model is invoked and the existing weights are loaded into the model. This is the first phase of the model. During this phase, we can see that the memory gradually increases by an amount of 430 MB. During the training phase, we can see that there is a further increase of 1.5 GB. This state of memory is continued throughout the phase of training. The next phase involves the saving of trained weights. This phase consumes a memory of 330 MB. To summarize, the training phase consumes the maximum memory out of all the phases.

\subsection{Training multiple accents}
We intend to evaluate our on-device training procedure on multiple accents. We conduct our experiment on the dataset for accents defined in Section \ref{sec:data acquisition}. Moving forward, we train the same $S_1$ model with batch size, B=5, and learning rate 10e-5, as these were the most effective hyperparameter values selected from our previous experiments.

Figure \ref{fig:NE_dataset results} shows the results, where the WER from the baseline model in blue and the WER from the saved model trained with stopping criteria in red.
The average WER before training is 25.11\%, and after training is 17.7\%. We see an average drop of 44\% for the WER across all accents. The results clearly show that on-device training improves the acoustic model for each speaker. 

Figure \ref{fig:ne_wertrend} shows the WER trend versus epochs for the first round of training. The trend indicates that WER values are monotonically decreasing with the increase in training epochs. The * symbol denotes the epoch where the minimum metric value (WER) or when the battery percentage or RAM is below the threshold is obtained, and the checkpoints are saved for the next round. We see that the steepest decrease in metric value is in the first epoch, and WER decreases slowly over the successive epochs. We do not observe any particular trend concerning accent or gender.

\subsection{Training for multiple rounds}
In this experiment, we aim to show the efficacy of on-device personalization by running multiple rounds. With this, we intend to show how the model adapts to a user's voice over time.
With the stopping criteria in place, we ensure that the model is not overfitting and has enough battery backup to carry out training. For this experiment, we consider four accents. We perform the second round of training using the model saved after using the stopping criteria in the first round.

\begin{figure}[!t]
	\centering
    \includegraphics[width = 0.5\textwidth]{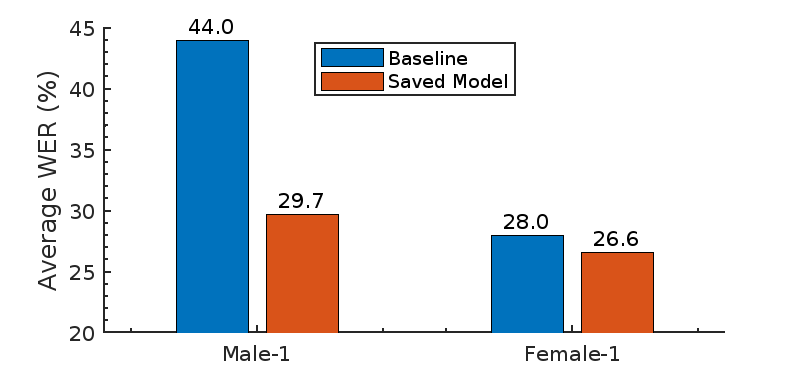}
    \caption{WER for real-time recorded voices}
    \label{fig:realwer}
\end{figure}
Figure \ref{fig:r2werne} shows the initial and saved model values of WER for the second round of training. We observe a reduction in WER for all accents. This confirms that the model is improving upon training for multiple rounds. We also present the WER trend for two training rounds in Figure \ref{fig:r2wertrend}. The trend of the first round is depicted in solid lines, and the second round is shown in dashed lines for all accents. The metric values gradually decrease in small increments as compared to round one. The majority of accents take greater than ten epochs to converge to minimum WER.

\subsection{Real-time recordings}

Finally, we validate our on-device training framework using real-time human speech recordings. We record samples from one male and a female subject with an Indian accent and train the baseline model. From Figure \ref{fig:realwer}, we can see a significant decrease in WER for the male voice. In contrast, the female voice does not show much improvement. The male subject is trained for eight epochs, and the female subject stopped at the first epoch.

\section{Summary}
\label{sec:summary}
In this paper, we present a methodology for implementing voice-personalized ASR models by training on multiple mobile phone brands. Our approach involves a resource-aware sub-model-based training method that considers mobile phones' limited RAM and battery capabilities.
Through our investigation of the relationship between available resources and training time, we highlight the effectiveness of using sub-models in such scenarios. By taking into account the evaluation metrics and battery constraints, we can perform efficient training and halt the process when necessary.
We created a speech dataset with multiple accents and trained the model with a single accent each time to simulate a real user. We use longer speech utterances of 7 to 12 seconds in length, with a maximum label length of 180 characters. Thus, we demonstrate the full functionality of our methodology by simulating a much more complex task than simple tasks such as speech command recognition, which typically are fixed-size shorter utterances. Furthermore, we provide insights into CPU and memory usage at various stages of training on a variety of mobile phones. 
Our system is well-suited for use in real-world federated learning ASR tasks as well as voice-controlled home automation applications that require locally trained ASR models.

\bibliographystyle{ACM-Reference-Format}
\bibliography{sample-base}

\newpage
\appendix
\section{Model}
\label{app:model}
\begin{table}[]
\centering
\caption{Total number of trainable parameters for different parts of the model.}
\label{finetuned-models}
\begin{tabular}[t]{lcc}
\toprule
    Layers&Number of trainable &Percentage\\
    & parameters &\\
\midrule
CONV 1-3&45.73k&0.14\\
BLSTM 1-4 &29.11M&96.26\\
FC 1-2 &1.08M&3.6\\
ALL &30.24M&100.0\\

\bottomrule
\end{tabular}
\end{table}%

\begin{table}[]
\centering
\caption{Partial models with the \% of parameters}
\label{tab:partial_models}
\begin{tabular}[t]{ccc}
\toprule
    Training Modes&Partial models&Percentage\\
\midrule
S1&CONV 1-FC 2&1.0\\
S2&CONV 2-FC 2&0.9999\\
S3&CONV 3-FC 2&0.9991\\
S4&BLSTM 1-FC 2&0.9985\\
S5&BLSTM 2-FC 2&0.6603\\
S6&BLSTM 3-FC 2&0.4521\\
S7&BLSTM 4-FC 2&0.2439\\
S8&FC 1-FC 2&0.0357\\
\bottomrule
\end{tabular}
\end{table}%
We use an end-to-end acoustic model with DeepSpeech2 (DS2) \cite{ds2} as the baseline model for ASR training. The architecture utilizes a well-optimized RNN based training system that does not require extensive pre-processing using phonemes to train.
The model consists of three convolutional layers (CONV 1-3) for feature extraction, four bi-directional long short term memory (BLSTM 1-4) layers with 1024 units in each direction, and two fully connected layers (FC 1-2) with 1024 units. The DS2 model has approximately 30.24M trainable parameters. Table \ref{tab:partial_models}
 shows the total number of parameters for different parts of the model.

\subsection{Partial models}
\label{app:sub_models}
In this section, we delve into the partial model-based training approach. By examining Table \ref{finetuned-models}, we observe that the number of parameters heavily relies on the layer type. Specifically, for the DS2 model, the BLSTM layers alone account for 96.26\% of the total parameters.

To investigate further, we generated partial models by progressively freezing the top layers and calculated the number of trainable parameters in each subpart of the model. Table \ref{tab:partial_models} presents all the possible partial models that can be constructed using this approach. We adopt a top-to-bottom methodology in creating these models, as the bottom layers have a more significant impact on the training process.

\begin{table}[!ht]
\centering
\caption{Speaker breakdown for on-device training dataset}
\begin{tabular}{|c|c|cc|cc|}
\hline
\multirow{2}{*}{\textbf{Datasets}}                                            & \multirow{2}{*}{\textbf{Accents}}                     & \multicolumn{2}{c|}{\textbf{\begin{tabular}[c]{@{}c@{}}Male\\ speakers\end{tabular}}}                                                     & \multicolumn{2}{c|}{\textbf{\begin{tabular}[c]{@{}c@{}}Female\\ speakers\end{tabular}}}                                                  \\ \cline{3-6} 

 &                                                     & \multicolumn{1}{c|}{\textbf{No.}}                                      & \textbf{\begin{tabular}[c]{@{}c@{}}Avg. len.\\ (s)\end{tabular}} & \multicolumn{1}{c|}{\textbf{No.}}                                      & \textbf{\begin{tabular}[c]{@{}c@{}}Avg. len\\ (s)\end{tabular}} \\ \hline
 
\multirow{1}{*}{\textbf{TTS}} & \begin{tabular}[c]{@{}c@{}}US\\ UK\\ IND\end{tabular} & \multicolumn{1}{c|}{\begin{tabular}[c]{@{}c@{}}1\\ 1\\ 1\end{tabular}} & \begin{tabular}[c]{@{}c@{}}9.3\\ 11.57\\ 11.96\end{tabular}      & \multicolumn{1}{c|}{\begin{tabular}[c]{@{}c@{}}1\\ 1\\ 1\end{tabular}} & \begin{tabular}[c]{@{}c@{}}10.0\\ 10.47\\ 11.6\end{tabular}     
\\ \cline{1-6} 
\textbf{\begin{tabular}[c]{@{}c@{}}Real-\\ time\end{tabular}}     & IND                                                  & \multicolumn{1}{c|}{1}                                                 & 10.2                                                             & \multicolumn{1}{c|}{1}                                                 & 10.7                                                            \\ \hline
\end{tabular}
\label{tab:speaker_breakdown}
\end{table}

\begin{table}[!ht]
\centering
	\caption{List of training datasets}
	\begin{tabular}[t]{lccc}
		\hline
		Dataset&Total Duration&Male&Female\\
		&(hrs)&(\%)&(\%)\\
		\hline
		LibriSpeech \cite{librispeech}&960&52&48\\
		Commonvoice \cite{commonvoice}&2000&45&15\\
		TEDlium \cite{tedlium}&250&66&34\\
		Fischer \cite{fischer}&2000&53&47\\
		\hline
	\end{tabular}
	\label{tab:datasets}
\end{table}

\begin{table*}[!ht]
	\centering
	\caption{Details about device hardware.}	
	\label{tab:different_phones}
	\begin{tabular}[t]{lcccccc}
		\hline
		\textbf{Device Model}&\textbf{RAM}&\textbf{CPU}&\textbf{OS}&\textbf{SoC (Snapdragon)}\\
		\hline
		OnePlus 7T&4.5|8 GB&Octa-core Max 2.96 GHz&Oxygen 10 (Android 10)&855 Plus\\
		OnePlus 5T&4|6 GB&Octa-core Max 2.45 GHz&Oxygen (Android 10 )&835\\
        Redmi Note 10 Pro Max&4.5|8 GB&Octa-core Max 2.3 GHz&MIUI V12.5.2(Android 11)&732G\\
		Redmi Note 7 Pro&2.8|6 GB&Octa-core Max 2.02 GHz&MIUI V12.5.1(Android 10)&675\\
		\hline
	\end{tabular}
\end{table*}%

\begin{figure*}[!ht]
	\centering
	\begin{subfigure}[b]{0.33\textwidth}
		\centering
		\includegraphics[width = 3.8cm,height= 7cm]{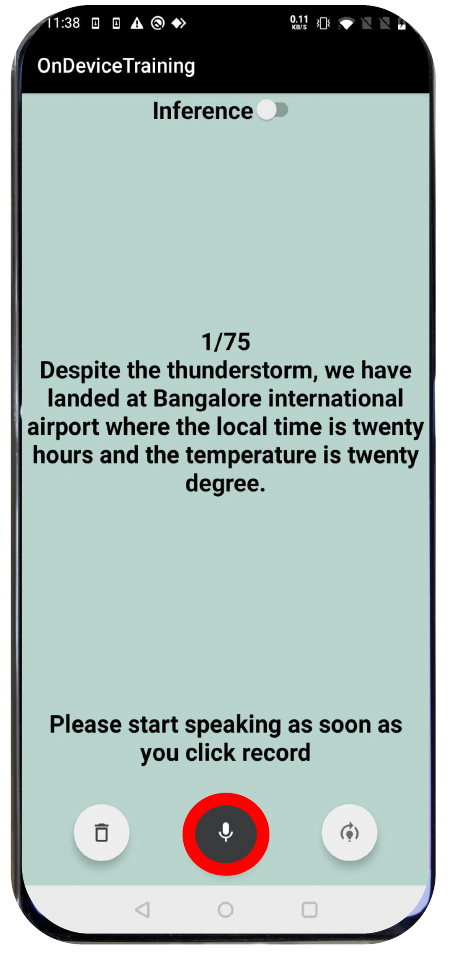}
		\caption{GUI for recording samples}
		\label{fig:recording}
	\end{subfigure}
	\begin{subfigure}[b]{0.33\textwidth}
		\centering
		\includegraphics[width = 3.8cm,height= 7cm]{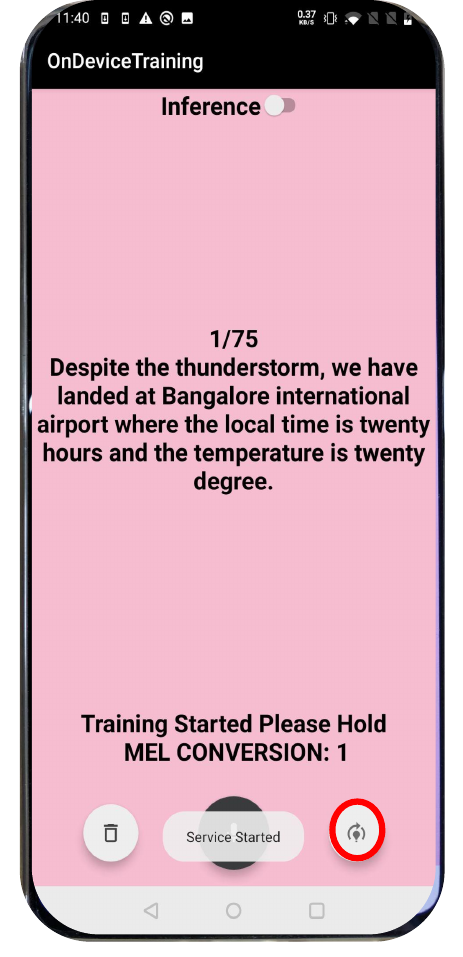}
		\caption{On-device training}
		\label{fig:training}
	\end{subfigure}
  	\begin{subfigure}[b]{0.33\textwidth}
		\centering
		\includegraphics[width = 3.8cm,height= 7cm]{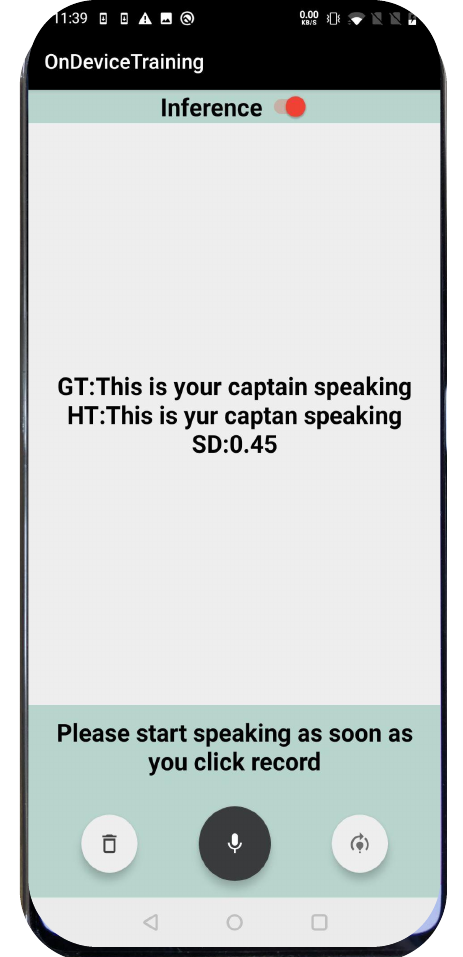}
		\caption{On-device inference}
		\label{fig:inference}
	\end{subfigure}
	\caption{Experimental setup}
	\label{fig:experimentalsetup}
\end{figure*}

\section{Datasets}
\label{app:datasets}

We created an audio corpus using a text-to-speech (TTS) system \cite{naturalreader} for 6 different accented speakers to simulate unique clients. The information regarding the various accents utilized and the average length of speech samples per accent is presented in Table \ref{tab:speaker_breakdown}. Each speech sample is about 9-12 seconds, with an average label length limited to 180 characters. We run the FL experiments by associating one speaker data to one client. The objective of this experiment to make the global model robust to multiple accents by learning from different accented clients. We use an end-to-end acoustic model similar to DeepSpeech2 \cite{ds2} architecture, collectively trained on datasets such as Librispeech \cite{librispeech}, commonvoice \cite{commonvoice} and tedlium \cite{tedlium} as our initial global model. 
The input speech samples are divided into windows of 32 milliseconds with 50\% overlap and converted to frequency domain. From each frame, a 80-dimensional log-melspectrogram is extracted to be given as input to the network.

\section{Mobile phone based Evaluation}
\label{app:andoid_app}

The application has a GUI to record samples for training procedure as shown in Figure  \ref{fig:recording}. Once we have enough data, we can train the underlying model and results will be displayed on the screen as shown in Figure  \ref{fig:training}. There is also a GUI to initiate only on-device inference and this is shown in Figure \ref{fig:inference}. 
The resource information from the mobile phone is collected using the aforementioned Android application.

The android application is installed on 4 mobile phones from three manufacturers with hardware specifications as mentioned in Table \ref{tab:different_phones}. We save the datasets for training and testing in the storage cache of the mobile phone. The application supports the user to either record the samples or use the dataset available in the cache. Table \ref{tab:different_phones} shows the hardware specifications of the multiple mobile phones considered in our work. We use four mobile phones with different resource capabilities to evaluate our approach. We considered phones with 6GB and 8GB RAM, and Snapdragon processors from 6xx series to 8xx series. The column related to RAM also shows the available memory.

\section{Estimating the memory footprint}
\label{app:memory_estimation}
In this section, we discuss the methodology we adopted to find the memory footprint of the model. 
To estimate the value of $\lambda$ for the model, we initially estimated $R_T$ and $R_L$. Here we show the calculation of $\lambda$ for S1. For determining $R_T$, we first find the number of parameters required to store size of the model ($n_m$), loss calculation ($n_l$), storing gradients ($n_g$), activations ($n_a$), errors ($n_e$) and optimizer parameters ($n_o$). For the DS2 architecture, the total number of parameters in the complete model is $n_m = 30,242,752 \approx 30.24M$. The total number of trainable parameters is only 30,240,574, and there are 2,178 non-trainable parameters in the model. The parameters to be stored during back-propagation including the number of gradients, errors per parameter and optimizer parameters depends only on the number of trainable parameters. Hence, we get $n_g=n_e=30,240,574$. For adaptive optimizer like Adam with momentum parameters this becomes $n_o=2 \times n_g$. We also find the size of the activations from each layer, which is required to be stored for calculating the gradients in back-propagation. We get $n_a = 68,816,40$, which is twice that of the model size $n_m$. Since it is difficult to find the memory requirement for a dynamic programming based loss function like CTC loss, where it consider many possible alignments between input and labels. We evaluated the memory requirement for this separately. For loss calculation, first we need to do forward propagation which need $n_m$, and $n_a$ parameters in memory, and then calculate loss for a mini-batch of 5 samples with an extra 100MB with CTC function, $\frac{(n_m+n_a*5)*4}{1024^2}+100 = 350MB$. 
Now we calculate the total size required for training a mini-batch of 5 samples, to get $N_T$. We also added an extra 200MB overhead for other parameters which is not accounted and system implementation. From this we calculated the memory required for training $R_T = 1.65GB$. Similarly, we calculated $R_L$ using $n_m$ with additional space for tensor information, and checkpoint size, and obtained $R_L \approx 325MB$.  Hence, we get $\lambda \approx 2GB$ for the complete model.

\begin{table}[]
\centering
\caption{Selected partial models S1, S5 and S8 (training modes) with the \% of parameters, and estimated $\lambda$ (GB)}
\label{tab:lambda_estimation}
\begin{tabular}[t]{ccc}
\toprule
    Training Modes&Percentage&Estimated $\lambda$\\
\midrule
S1&1.0&1.98\\
S5&0.6603&1.63\\
S8&0.0357&0.96\\
\bottomrule
\end{tabular}
\end{table}%
Similarly, we calculated $\lambda$ for all the selected sub-models as shown in the Table \ref{tab:lambda_estimation}, where the number of trainable parameters vary according to the layers selected.


\end{document}